\documentclass[epj]{svjour}
\usepackage{graphics}
\begin{document}
\title{Models of random graph hierarchies}
\author{Robert Paluch\inst{1} \and Krzysztof Suchecki\inst{1} \and Janusz A. Ho\l yst \inst{1,2}
}                     
%
%
\institute{Center of Excellence for Complex Systems Research, Faculty of Physics, Warsaw University of Technology,\\Koszykowa 75, PL-00662 Warsaw, Poland \and ITMO University, 19 Kronverkskiy av., 197101 Saint Petersburg, Russia}
\date{Received: date / Revised version: date}
%
\abstract{
We introduce two models of inclusion hierarchies: Random Graph Hierarchy (RGH) and Limited Random Graph Hierarchy (LRGH).
In both models a set of nodes at a given hierarchy level is connected randomly, as in the Erd\H{o}s-R\'{e}nyi random graph, with a fixed average degree equal to a system parameter $c$.
Clusters of the resulting network are treated as nodes at the next hierarchy level and they are connected again at this level and so on, until the process cannot continue.
In the RGH model we use all clusters, including those of size $1$, when building the next hierarchy level, while in the LRGH model clusters of size $1$ stop participating in further steps.
We find that in both models the number of nodes at a given hierarchy level $h$ decreases approximately exponentially with $h$.
The height of the hierarchy $H$, i.e. the number of all hierarchy levels, increases logarithmically with the system size $N$, i.e. with the number of nodes at the first level.
The height $H$ decreases monotonically with the connectivity parameter $c$ in the RGH model and it reaches a maximum for a certain $c_{max}$ in the LRGH model.
The distribution of separate cluster sizes in the LRGH model is a power law with an exponent about $-1.25$.
The above results follow from approximate analytical calculations and have been confirmed by numerical simulations.
}

\maketitle
\section{Introduction}
\label{intro}
Hierarchical structures can be found in many forms in many real world systems.
Four general classes of hierarchical systems can be distinguished \cite{pumain} -- order, control, inclusion and level hierarchies.
An {\it order} hierarchy is a set of units ordered by an internal variable attributed to these units, e.g. a company income, a book size or a simple social rank \cite{bonabeau}.
{\it Control} hierarchies describe control relations such as boss-subordinate \cite{lopez} or leader-followers \cite{hierarchygrowth} and are usually represented by directed graphs \cite{mones,corominas}.
{\it Inclusion} hierarchies \cite{ravasz,logperiodic,clauset} correspond to stuctures where a unit of a higher level includes several units of a lower level, e.g. a university includes faculties or an army division includes regiments.
{\it Level} hierarchies can be treated as a special class of inclusion hierarchies when interacting elements of a lower level collectively form elements of a higher level and the higher level elements possess some emergent properties absent at the lower level \cite{brain}.
Examples are cells forming tissues and organs or macromolecules forming internal cell structures.
The concept of inclusion and level hierarchies is close to the community structure of networks \cite{newman,fortunato_community} that is defined through a network topology, and to recently studied networks of networks \cite{interdependent,interconnected_spreading,reis}
The original community structure measures considered a simple division of nodes into disjoint communities.
This has changed, as more recent works focus on overlapping communities, that are able to form an inclusion hierarchy \cite{clauset,xie}.
In fact, Clauset \cite{clauset} introduces a very general model of a hierarchical graph.
His model is descriptive in nature, as it does not determine connection probabilities across the hierarchy.
Hierarchical community organization has been also considered for several dynamical models \cite{galam,scirep,agnieszka2}.
In this paper, we introduce two models of inclusion hierarchies, with a specified mechanism of hierarchy organization.
The models are not strictly network models but can be interpreted in terms of nested community structures, or in terms of networks of networks featuring a nested hierarchical topology. 
While the first investigated model is straightforward, the second one shows a nontrivial dependence of the total number of levels on system parameters.
In the following sections we define the RGH and LRGH models and derive their basic properties, mainly the total number of hierarchy levels that exist in the system.

\section{Random Graph Hierarchy} 
\label{HER}
We propose a hierarchical system consisting of $N$ nodes which are organized into many nested clusters.
Our idea is to recursively repeat the connection procedure of the well-known Erd\H{o}s-R\'{e}nyi graph and treat connected clusters created at each level as nodes at a next level.
The algorithm for creating such structure is following:
\begin{enumerate}
	\item Link together each pair of nodes with a probability $p_0=c/(N-1)$. The result is $W_1$ clusters of the hierarchical degree $h=1$. The nodes belonging to the same cluster are considered as neighbours, so their mutual degree of neighbourhood is one.
	\item The existing clusters of hierarchical degree $h=1$ are treated as primary-level nodes which are linked together randomly with a probability $p_1=c/(W_1-1)$ per pair. The result is $W_2$ clusters of the hierarchical degree $h=2$, containing sets of clusters of degree $h=1$. If nodes $i$ and $j$ belong to the same cluster with the hierarchical degree $h=2$, but to different clusters of the hierarchical degree $h=1$, their mutual degree of neighbourhood is two.
	\item Repeat step 2 for next hierarchy levels $h=3,4,5 \dots$ until only one cluster remains. The total number of iterations $H$ required is proportional to the logarithm of the system size $H \sim \ln N$.
\end{enumerate}
The model possesses only two parameters: $N$ -- the total number of nodes and $c$ -- the mean number of connections per node created at each level.
Note that making the parameter $c$ constant instead of connection probability $p$ known from E-R model means that despite a changing number of clusters $W_h$, the connection density is always the same in relation to the percolation threshold ($\langle k \rangle =1$ in random graphs) at each level $h$s.
The resulting structure will be called {\it Random Graph Hierarchy} (RGH).
The average number of clusters $W_h$ of hierarchical degree $h$ decreases as $W_h = N \alpha^h$, where $\alpha = \alpha(c) = 1 - c/2$.
This is due to the average number of clusters in a random graph of $N$ nodes equal to $n_c = N(1-c/2) + O(1)$ (this is true only below the percolation threshold, i.e. for $c<1$) \cite{cont_bouchaud}.

\begin{figure}[htb]
\resizebox{1\columnwidth}{!}{\includegraphics{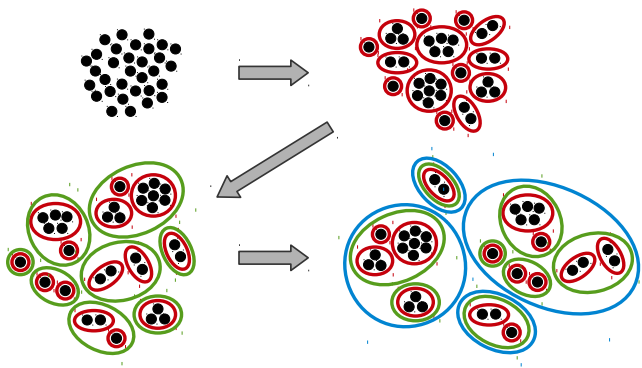}}
\caption{The building process of the Random Graph Hierarchy (RGH). During each step nodes/clusters are randomly connected (as in the Erd\H{o}s-R\'{e}nyi random graph) to form clusters. The clusters are connected in the next step in the same way and so on. The process continues until the whole system is aggregated in a single cluster. The coloured ellipses show the clusters (red for level 1, green for level 2 and blue for level 3).}
\label{Fig:scheme1}
\end{figure}

The total number of hierarchies $H$ fulfills the following equation:
\begin{equation}
W_H = N (1-\frac{c}{2})^H = 1.
\label{eqWH}
\end{equation}
It follows that
\begin{equation}
H = - \frac{\ln N}{\ln(1-c/2)}.
\label{eqH1}
\end{equation}
Eq. \ref{eqH1} is only an approximation since it is based on the average number of clusters in E-R graph which is correct only for large $N$ what is not fulfilled in the case of $h$ close to $H$.

\begin{figure}[htb]
\resizebox{1\columnwidth}{!}{\includegraphics{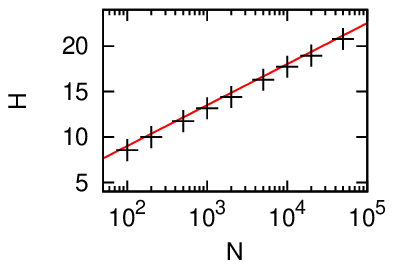} \hskip -0.66cm \includegraphics{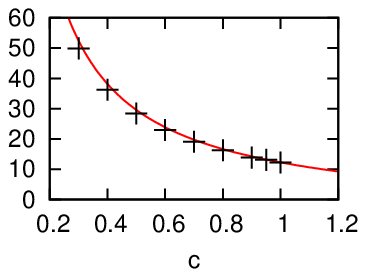}}
\caption{The highest level of hierarchy $H$ as a function of the system $N$ (left picture, $c=0.8$, averaged over $500$ realizations) and average degree $c$ (right picture, $N=5000$, averaged over $500$ realizations) for RGH model. The lines are analytical results (Eq. \ref{eqH1}). Error bars are smaller than symbol sizes.}
\label{Fig:hierarchies1}
\end{figure}

Note that while the resulting structure is a graph at each hierarchy level, it does not possess well-defined links as a whole, since higher-level links are between clusters, not between individual nodes.

\section{Limited Random Graph Hierarchy}
In the RGH model each node/cluster is given many chances to join other elements.
Even if an element of level $h$ does not connect with any other, it advances to level $h+1$ as a cluster of the size one.
We present an alternative approach in the {\it Limited Random Graph Hierarchy} (LRGH) where {\it only} clusters that have merged at the level $h$ can participate in the merging at the level $h+1$.
Otherwise they drop out and do not participate in further merging.
In effect the procedure of the cluster growth is \emph{limited} only to the groups which are continually developing.
Aside from dropping out nodes/clusters that do not find partners, the procedure is the same as for the RGH model and has the same two parameters: $N$ and $c$.
The procedure is as follows:
\begin{enumerate}
	\item Link together each pair of nodes with the probability $p_0=c/(N-1)$. The result is $W_1$ clusters (of size at least two) which advance to the second step. Unlinked nodes are the clusters of the size one and they do not participate in the further steps (and they are not included in $W_1$).
	\item Merge together each pair of clusters with the probability  $p_1=c/(W_1-1)$. The result of merging two clusters is a new cluster which contains these sub-clusters and whose size is the sum of the sizes of the merged clusters. The new clusters ($W_2$) advance to the next step. The clusters which did not merge during this step do not advance and stop growing (and they are not included in $W_2$).
	\item Repeat the step 2 for the clusters of successive levels until all clusters stop growing.
\end{enumerate}

\begin{figure}[htb]
\resizebox{1\columnwidth}{!}{\includegraphics{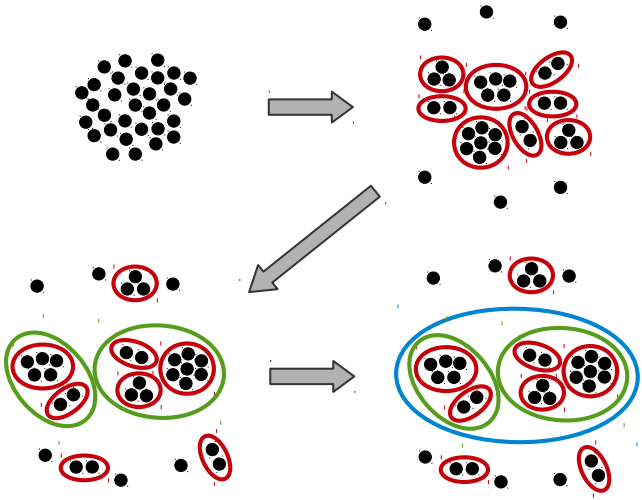}}
\caption{Multilevel growth of clusters in the LRGH model. The nodes/clusters at every level are randomly connected (as in Erd\H{o}s-R\'{e}nyi random graph) to form clusters. The nodes/clusters which did not connect and therefore remain alone become inactive and are exempt from further dynamics. The clusters formed from the successful connections remain active and are connected in the next step in the same way. The process continues until no active clusters remain. The coloured ellipses show the clusters (red for level 1, green for level 2 and blue for level 3).}
\label{Fig:scheme2}
\end{figure}

Similarly to RGH model, the number of clusters on given hierarchy level $W_h$ and the maximum hierarchy level $H$ can be calculated analytically.
At the beginning we find the average number of clusters $W_h$ of level $h$ (which advanced to $h+1$-th step, obviously $W_0 = N$).
Consider the first step of the above procedure.
As previously noted in Sect. \ref{HER}, the average number of clusters resulting from the linking together each pair of $W_0$ nodes with the probability $p_0=c/(W_0-1)$ is $n_0 = W_0(1-c/2)$, including clusters of the size one (single, unlinked nodes).
The average number of unlinked nodes $R_0$ is given by the probability of $W_0-1$ failures to link multiplied by the number of all nodes:
\begin{equation}
R_0 = W_0\left(1-\frac{c}{W_0-1}\right)^{W_0-1},
\end{equation}
therefore the average number of clusters $W_1$ which advanced is the difference between $n_0$ and $R_0$:
\begin{equation}
W_1 = n_0 - R_0 = W_0\left[1 - \frac{c}{2} - \left(1-\frac{c}{W_0-1}\right)^{W_0-1}\right],
\end{equation}
leading to a recursive equation for $W_h$:
\begin{equation}
W_h = W_{h-1}\left[1 - \frac{c}{2} - \left(1-\frac{c}{W_{h-1}-1}\right)^{W_{h-1}-1}\right].
\label{eqWh_rec}
\end{equation}
Since Eq. \ref{eqWh_rec} does not depend explicitly on $h$, we assume that $W_{h-1}$ is large enough to use the approximation
\begin{equation}
(1-\frac{c}{W_{h-1}-1})^{W_{h-1}-1} \approx e^{-c},
\label{eqApprox}
\end{equation}
which is independent from $h$ and $W_h$.
The accuracy of this approximation depends mainly on $W_{h-1}$, the error not exceeding $10\%$ when $W_{h-1} \geq 7$.
In practice, for all lower hierarchy levels with their numerous clusters, it is a very good approximation, while introducing some error at the top level.
Since the number of clusters at the last level $H-1$, where connections still appear, is around $7$ or higher, the error of Eq.\ref{eqApprox} is at most $10\%$, but it causes upwards to $20\%$ discrepancy in the following Eq. \ref{eqWh}.
Note that this error causes a systematic underestimation of $W_H$ (and in consequence a corresponding underestimation of $H$).
The error is smaller for smaller $c$ and is independent of the system size $N$ and the number of hierarchy levels $H$.

\begin{figure}[htb]
\resizebox{1\columnwidth}{!}{\includegraphics{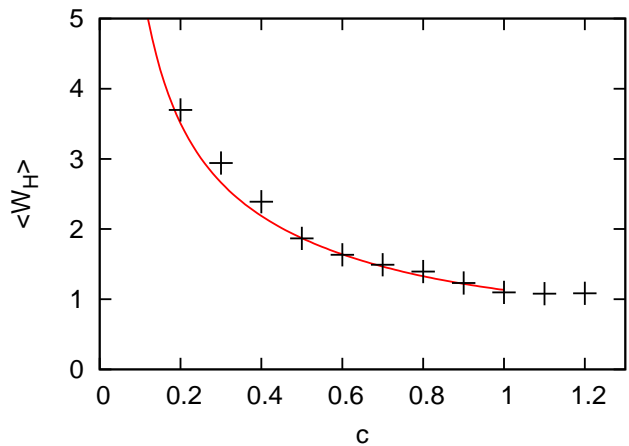}}
\caption{The average number of clusters observed at the final level $H$. The crosses are numerical values averaged over $2000$ realizations ($N=5000$). The line is an analytical result calculated using Eq. \ref{eqW_H}. Error bars are smaller than symbol sizes.}
\label{Fig:WHc}
\end{figure}

Using the approximation (Eq. \ref{eqApprox}) we obtain
\begin{equation}
W_h = N\left(1 - \frac{c}{2} - \mathrm{e}^{-c}\right)^h,
\label{eqWh}
\end{equation}
which describes the dependence of number of active clusters on the hierarchy level $h$.
We can invert this equation, calculating the hierarchy level $h$ of a system with $W_h$ clusters.
Then, using it for $h=H$ we can write
\begin{equation}
H(c,N) = \frac{\ln\left( \langle W_H \rangle / N \right)}{\ln\left(1-\frac{c}{2}-\mathrm{e}^{-c}\right)}.
\label{eqH2}
\end{equation}
Let us first assume that $\langle W_H \rangle = 1$, meaning that the last hierarchy always consists of a single cluster merged in the previous step.
If we do so, then Eq. \ref{eqH2} depends on the value of $c$ only through the argument of the logarithm, and possesses a maximum for the argument $c_{max}=\ln 2 \approx 0.6931$.
While this prediction reproduces the general shape of the relation correctly, both the values and the position of the maximum are underestimated.
Using a different terminating constant than $\langle W_H \rangle = 1$ changes the height, but the position of the maximum remains the same.\\
To get a better approximation, it is necessary to find a more precise value of $\langle W_H \rangle$.
The procedure of LRGH creation stops when none of the existing clusters connects with any other.
The probability that it happens depends on the number of participating clusters $W$ and the constant $c$:
\begin{equation}
P_c(W,c) = \left(1 - \frac{c}{W-1}\right)^{\frac{W(W-1)}{2}}.
\end{equation}
This probability is conditional on the number of clusters $W$ present in the system.
The probability that the system passes through a state with a given number of clusters $W$ is equal to the probability that it has not finished the dynamics at earlier steps.
However, since the system does not visit each number of clusters $W$ in order, but visits only certain values during a single realization, we approximate it by treating it as if the system visits all of them, but spends less than one step in each.
Effectively we consider $h$ as the time, since one procedure step changes it by $1$ each time.
Since the system spends less time in each $W$, the probability $P_c$ of terminating the process should be multiplied by the time spent at a given value of $W$.
It can be approximated as the inverse of the absolute value of derivative $\mathrm{d}W/\mathrm{d}h$.
Using Eq. \ref{eqWh} for $W(h)$, we can write the conditional probability of terminating the process with $W$ clusters scaled by the time spent with $W$ clusters during the evolution
\begin{equation}
P_{ct}(W,c)=\left| \frac{\mathrm{d}h}{\mathrm{d}W} \right| P_c(W,c)= \frac{-P_c(W,c)}{W \ln (1-c/2-\mathrm{e}^{-c})}.
\end{equation}
Now the probability to complete the dynamics at a given number of clusters $W$ is
\begin{equation}
P_a(W,c)=P_{ct}(W,c) \prod_{k=W+1}^{N} (1-P_{ct}(k,c)).
\label{eqpabs}
\end{equation}
Unfortunately this probability is not normalized, since our approach does not describe the behaviour of the system correctly after it arrives at $W=1$.
We therefore treat it as the probability to terminate before $W=1$ (it is always lower than $1$).
If the system manages to merge down to one cluster, then the process stops, as any further merging is impossible.
We therefore assume that the missing probability is the chance to stop at $W=1$.
This leads to the final formula for the average number of clusters at the hierarchy $H$
\begin{equation}
\langle W_H \rangle = \sum_{W=2}^{N} W P_a(W,c) + (1-\sum_{W=2}^{N} P_a(W,c)),
\label{eqW_H}
\end{equation}
where the first term is average of $W$ over probabilities $P_a(W,c)$ for $W>1$, while the second term is the probability to stop at $W=1$.
Because of the high algebraic complexity, we have not managed to express Eq. \ref{eqW_H} in a simple analytical form, and instead we have calculated this value numerically. 
We have observed that while the products and series presented by Eq. \ref{eqW_H} and Eq. \ref{eqpabs} formally depend on $N$, in practice the terms quickly become very small for large $W$ and $\langle W_H \rangle$ attains a limit value independent of $N$.
We used only the first $100$ terms (as if $N=100$) in our study, which for $c>0.1$ produced an error no larger than $0.05$ that quickly decreases with $c$ (for $c=0.5$ it is less than $10^{-10}$).
Figure \ref{Fig:WHc} shows that despite many approximations and assumptions used, the value of $\langle W_H \rangle$ calculated this way is very close to what is observed during numerical simulations.


\begin{figure}[htb]
\resizebox{1\columnwidth}{!}{\includegraphics{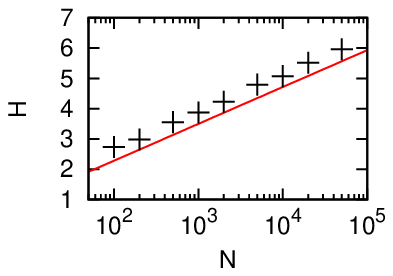} \hskip -0.66cm \includegraphics{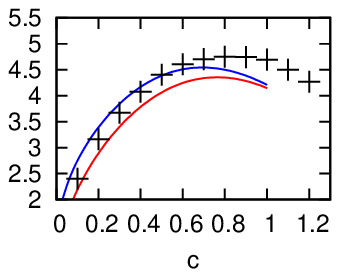}}
\caption{The highest level of hierarchy $H$ as a function of the system size $N$ (left picture, $c=0.8$, averaged over $500$ realizations) and the connectivity $c$ (right picture, $N=5000$, averaged over $2000$ realizations) for LRGH model. The red lines are analytical results (combined Eq. \ref{eqH2} and Eq. \ref{eqW_H}), while the blue line (right picture) shows Eq. \ref{eqH2} under assumption $\langle W_H \rangle = 1$. Error bars are smaller than symbol sizes.}
\label{Fig:hierarchies2}
\end{figure}

Putting Eq. \ref{eqW_H} into Eq. \ref{eqH2}, we obtain the value $H(c,N)$ that reproduces the numerical results with a greater accuracy than the value $H(c,N)$ based on the assumption $\langle W_H \rangle = 1$, as seen in Fig. \ref{Fig:hierarchies2}.
However the discrepancies with the numerical results are still visible.
The numerical maximum of $H(c)$ is slightly shifted towards $c=1$, i.e. for $N=5000$ it is about $0.8$, while the analytical value $c_{max}(N=5000)=0.7631$.
These differences originate from two sources: we take the average value of $W_H$ (although this is a very good approximation, as seen in Fig. \ref{Fig:WHc}) and we use the approximation shown in Eq. \ref{eqApprox} which introduces a substantial error for the last hierarchies.
The position of the maximum $c_{max}$, due to $\langle W_H \rangle$ depending on $c$, now depends weakly on the system size $N$ (i.e. $c_{max}(N=100)=0.8178$).


\begin{figure}[htb]
\resizebox{1\columnwidth}{!}{\includegraphics{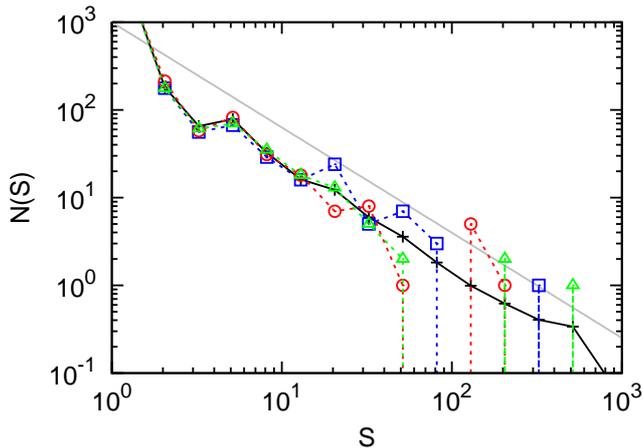}}
\caption{The distribution of cluster sizes for LRGH model ($c=0.8$, $N=5000$). Circles, triangles and squares are examples of single realizations. The black solid line shows the distribution averaged over $1000$ realizations. The gray straight line is a power-law with the exponent $\alpha=-1.25$, just for reference.}
\label{Fig:clustersizes}
\end{figure}

Unlike in the RGH model, where the clusters merge until only a single one is left at the level $H$, in the LRGH model clusters stop participating in further merging at the moment when they fail to form any connection in a given step.
This means that the final structure consists of many disconnected clusters.
We investigated the distribution of these clusters and observed a dependence similar to a power law (Fig. \ref{Fig:clustersizes}). 
Due to the a relatively small exponent ($\alpha \approx -1.25$) in a single realization of the system the power-law tail is disrupted by existence of the largest cluster.
This behavior is seen in Fig. \ref{Fig:clustersizes}, where single realizations feature one or few very large clusters that may dominate the whole system, and the peaks corresponding to these clusters are often clearly separated from the quasi-continuous part of the distribution.

\section{Conclusions}


We have introduced two models of hierarchical structures, based on random connections between the nodes (as in Erd\H{o}s-R\'{e}nyi graphs) at different hierarchy levels $h$, with a constant mean node degree independent from $h$.
Both models show an approximately exponential decrease of the number of elements at successive levels, and possess a well defined maximum hierarchy level $H$ (the height of the hierarchy).
The height $H$ is proportional to the logarithm of the number of nodes $N$.
It decays monotonicaly with the connectivity parameter $c$ in the Random Graph Hierarchy model, but possesses a maximum as a function of this parameter for the Limited Random Graph Hierarchy model.
The obtained analytical results are supported by numerical simulations.\\
The models can be used for investigation of opinion dynamics or other collective processes \cite{paluch_cb} on hierarchical structures.
Since many real social systems are hierarchical \cite{pumain} the models may offer a simple representation for them. 
The proposed models could be also considered as reference models for investigations of inclusion hierarchies \cite{pumain,ravasz,logperiodic,clauset}. 
We expect they may be useful for cases where a system self-organizes into a hierarchy, example being countries, which can arise from settlements joining into progressively larger regional powers or corporations arising from progressive mergers of smaller companies.
Our models generate structures that possess several levels of hierarchical organization.
While some real systems are shallow, with only one or two levels, in many of them several levels could be distinguished, such as labs, divisions, faculties and finally universities or townships, counties, individual states and the whole of United States.
Of particular interest may be the fact that in the LRGH model the final system can consist of clusters that stopped growing at various levels.
They can represent different organizational complexity, similar to countries that depending on their size (but not only) could have different depths of organization, however they are usually still considered same-level entities.

\section*{Acknowledgments}
The research leading to these results has received funding from the European Union Seventh Framework Programme
(FP7/2007-2013) under Grant Agreement No. 317534 (the Sophocles project) and from the Polish Ministry of Science and Higher Education Grant No. 2746/7.PR/2013/2. J.A.H. has been also partially supported by Russian Scientific Foundation, proposal \#14-21-00137  and by European Union COST TD1210 {\it KNOWeSCAPE} Action.

\bibliographystyle{spphys}
\bibliography{Paluch_HERG_bibliography}

\end{document}